\documentclass[aps, prd, floatfix, nofootinbib, superscriptaddress, twocolumn]{revtex4-2}

\usepackage{amsmath,amssymb,amsfonts}
\usepackage{graphicx}
\usepackage{bm}
\usepackage{booktabs}
\usepackage{multirow}
\usepackage{float}
\usepackage{xcolor}
\usepackage{adjustbox}
\usepackage[mathscr]{euscript}
\usepackage{dsfont}

\allowdisplaybreaks[2]

\usepackage[colorlinks=true,linkcolor=blue,citecolor=blue,urlcolor=blue]{hyperref}

\newcommand{\dd}{\mathrm{d}}

\begin{document}

\title{Electromagnetic Structure of $\Omega_{ccc}^{++}$ and
$\Omega_{bbb}^{-}$ baryons from a covariant three-body Faddeev equation}

\author{Zhao-Qian\ Yao%
$\,^{\href{https://orcid.org/0000-0002-9621-6994}{\textcolor[rgb]{0.00,1,0.00}{\sf ID}}}$}
\email{z.yao@hzdr.de}
\affiliation{\href{https://ror.org/01zy2cs03}{Helmholtz-Zentrum Dresden-Rossendorf},
Bautzner Landstra{\ss}e 400, D-01328 Dresden, Germany}

\begin{abstract}
We investigate the electromagnetic structure of the ground-state
$\Omega_{ccc}^{++}$ and $\Omega_{bbb}^{-}$ baryons using a
Poincar\'e-covariant three-body Faddeev equation within the
rainbow--ladder truncation.  We calculate the complete set of elastic
spin-$3/2$ multipole form factors,
$G_{E0}$, $G_{M1}$, $G_{E2}$, and $G_{M3}$.
The leading electric and magnetic form factors become markedly harder
from the charm to the bottom sector, revealing the substantially more
compact electromagnetic structure of the $\Omega_{bbb}^{-}$.
For $\Omega_{ccc}^{++}$, the electric form factor and charge radius at
low momentum transfer agree well with the available lQCD results.
From the magnetic-dipole form factor at zero momentum transfer,
$G_{M1}(0)$, we obtain the magnetic moments
$\mu_{\Omega_{ccc}^{++}}=1.26\,\mu_N$ and
$\mu_{\Omega_{bbb}^{-}}=-0.207\,\mu_N$.
For $G_{E2}$ and $G_{M3}$, the electric-quadrupole structure exhibits a stronger charm-to-bottom
evolution than the charge-normalized magnetic-octupole structure.
Finally, comparison of the charm and bottom systems shows that both
$G_{M1}(0)$ and the relation between the electric and magnetic radii
move closer to their leading nonrelativistic heavy-quark expectations
as the quark mass increases.

\end{abstract}
 % and a quark--photon vertex obtained from the corresponding inhomogeneous Bethe--Salpeter equation. We obtain masses of \(4.76\) and \(14.35\,{\rm GeV}\), respectively, in good agreement with available lQCD results. The \(\Omega_{bbb}^{-}\) Faddeev amplitude is harder in momentum space, consistent with its more compact spatial structure.

\maketitle
% ============================================================
\section{Introduction}
\label{sec:intro}
The observation of the doubly charmed baryon
$\Xi_{cc}^{++}(ccu)$ by the LHCb Collaboration marked an important milestone in heavy-baryon spectroscopy, demonstrating that baryons containing more than one charm quark are experimentally accessible at
hadron colliders~\cite{Aaij:2017ueg}. More recently, LHCb reported the
observations of the remaining doubly charmed baryons,
$\Xi_{cc}^{+}(ccd)$ and $\Omega_{cc}^{+}(ccs)$, thereby substantially
expanding the experimentally established doubly charmed baryon
spectrum~\cite{LHCb:2026pxn,LHCb:2026lnj}. These developments
naturally strengthen the prospect of extending heavy-baryon
spectroscopy to the triply-charmed sector. 
% Indeed, dedicated studies
% of triply-heavy baryon production have suggested that the production
% of $\Omega_{ccc}^{++}$ at the LHC may be sufficiently sizable to make
%its observation feasible~\cite{ChenWu:2011}.
Indeed, dedicated studies of triply-heavy baryon production,
including recent fragmentation-based analyses, have explored the
prospects for observing these states at the LHC and future
hadron colliders~\cite{ChenWu:2011,Celiberto:2025ogy,Celiberto:2026qiz}.
It is therefore
reasonable to anticipate that the first experimental information on a
triply-charmed baryon may become available in the foreseeable future.

The $\Omega_{ccc}^{++}$ and $\Omega_{bbb}^{-}$ are the simplest baryonic
systems in which three identical heavy quarks are bound by QCD~\cite{Meinel:2012qz,Padmanath:2013zfa,Qin:2019hgk}.
For a color-singlet ground state composed of three identical quarks,
the Pauli principle imposes particularly restrictive symmetry
constraints. Since the color wave function is antisymmetric, while the
flavor and spatial wave functions of the ground state are symmetric,
the spin wave function must also be fully symmetric. Consequently, an
$S$-wave $ccc$ or $bbb$ state with $J^{P}=1/2^{+}$ is forbidden, and
the lowest-lying $\Omega_{ccc}^{++}$ and $\Omega_{bbb}^{-}$ baryons
necessarily carry $J^{P}=3/2^{+}$
\cite{Meinel:2012qz,Padmanath:2013zfa}.

These systems therefore provide especially clean environments for
studying the internal structure of baryons composed entirely of heavy
quarks. In particular, their electromagnetic form factors encode the
spatial distribution of the heavy quarks and provide insight into the
dynamics responsible for binding them into a three-quark state.
Studying both the charm and bottom sectors further makes it possible to
trace how these structural properties change as the heavy-quark mass
increases, thereby providing a useful connection between relativistic
bound-state dynamics and the heavy-quark regime.

Continuum Schwinger methods (CSMs), formulated in terms of Dyson--Schwinger
equations (DSEs), provide a complementary framework for addressing this
problem~\cite{Roberts:1994dr,Maris:2003vk,Holl:2005st,Qin:2020jig,Cheng:2025yij,Xu:2025cyj,Xing:2023pms,Cheng:2022jxe}.  The quark gap equation, together with the Bethe--Salpeter and
Faddeev equations, enables a Poincar\'e-covariant and
symmetry-preserving treatment of mesons and baryons within a common
underlying interaction.  In particular, the covariant three-body
Faddeev equation has been successfully applied to studies of the masses
and electromagnetic properties of baryons~\cite{Wang:2018kto,Eichmann:2009qa,Yao:2024ixu,Qin:2018dqp,Qin:2019hgk,Eichmann:2011vu}.
Its extension to the triply-heavy sector therefore provides a natural
framework for investigating the electromagnetic structure of
$\Omega_{ccc}^{++}$ and $\Omega_{bbb}^{-}$, including not only the
electric-monopole and magnetic-dipole form factors,
$G_{E0}$ and $G_{M1}$, but also the higher electric-quadrupole and
magnetic-octupole form factors, $G_{E2}$ and $G_{M3}$. Together,
these form factors provide a comprehensive characterization of the
electromagnetic properties of spin-$3/2$ baryons.

These electromagnetic properties can be further examined from the
perspective of the heavy-quark limit.  Heavy-quark symmetries have been
employed in studies of doubly-heavy baryons~\cite{Liu:2018euh,Shi:2021kmm}.  A related
flavor evolution is also evident in the meson sector: continuum studies
of pseudoscalar and vector mesons show that their charge distributions
become increasingly localized with increasing current-quark mass, while
differences between spin-partner radii are progressively reduced toward
the heavy-quark regime~\cite{Xu:2024vkn,Xu:2026rtu}.
% These observations motivate us to investigate whether analogous
% heavy-quark signatures emerge in the electromagnetic structure of
% triply-heavy baryons.

In this work, we investigate the electromagnetic structure of the
ground-state $\Omega_{ccc}^{++}$ and $\Omega_{bbb}^{-}$ baryons using
the Poincar\'e-covariant three-body Faddeev equation in the
rainbow--ladder truncation.  A common interaction is employed to
determine the dressed heavy-quark propagators and the corresponding
Faddeev amplitudes.  We calculate all four elastic spin-$3/2$
multipole form factors, $G_{E0}$, $G_{M1}$, $G_{E2}$, and
$G_{M3}$, and extract the associated static electromagnetic
properties.  
% Particular attention is paid to the analytic continuation
% of the Faddeev amplitudes required at nonzero momentum transfer and to
% the numerical extraction of the higher multipoles near $Q^2=0$.

Beyond providing predictions for the triply-bottom system, the combined
analysis of $\Omega_{ccc}^{++}$ and $\Omega_{bbb}^{-}$ allows us to
examine the evolution of their electromagnetic structure with increasing
heavy-quark mass.  In particular, we study the approach of
$G_{M1}(0)$ to its leading nonrelativistic (NR) reference value and the
convergence of the electric and magnetic radii.  We find that the
bottom system is substantially more compact, while the magnetic moment
and the increasingly similar electric and magnetic radii both exhibit
a clearer approach toward heavy-quark behavior.
% ============================================================
\section{Hadronic Bound-State Equations}
In quantum field theory, a hadron is identified with a pole in the
appropriate even-point Green function, or equivalently in the
corresponding scattering matrix.  In Euclidean space, the mass shell
of a bound state with mass $M$ is located at
\begin{equation}
P^2=-M^2,
\end{equation}
where $P$ denotes the total momentum of the system.  In the vicinity
of the bound-state pole, the scattering matrix factorizes as
\begin{equation}
T \xrightarrow{\,P^{2}\to -M^{2}\,}
\frac{\Psi\,\bar{\Psi}}{P^{2}+M^{2}}
+\text{regular terms},
\label{eq:bound_state_pole}
\end{equation}
where $\Psi$ denotes the bound-state amplitude and $\bar{\Psi}$ its
conjugate amplitude.  The factorization of the pole residue expresses
the propagation of an intermediate hadronic bound state.

Inserting Eq.~\eqref{eq:bound_state_pole} into the corresponding
scattering equation and equating the pole residues yields the
homogeneous bound-state equation
\begin{equation}
\Psi = K\,G_0\,\Psi ,
\label{eq:homogeneous_bound_state}
\end{equation}
where $K$ is the interaction kernel and $G_0$ denotes the product of
the dressed propagators of the constituents.  For a quark--antiquark
meson, Eq.~\eqref{eq:homogeneous_bound_state} becomes the homogeneous
Bethe--Salpeter equation, whereas for a three-quark baryon it leads to
the corresponding homogeneous Faddeev equation.

\subsection{Gap equation and Bethe--Salpeter equations}
\label{sec:gap_bse}

In order to determine the hadron bound-state amplitudes, the dressed-quark
propagators are required.  They are obtained from the QCD gap equation~\cite{Roberts:1994dr}:
\begin{subequations}
\label{EqGap}
\begin{align}
S^{-1}(k) &= i\gamma\cdot k + m + \Sigma(k) \,,
\\
\Sigma(k) &=
\int\frac{\dd^4 q}{(2\pi)^4}
\,4\pi\mathcal{G}_{\mu\nu}(l)
\,\gamma_\mu\frac{\lambda^{a}}{2} S(q)
\Gamma_\nu(q,k)\frac{\lambda^{a}}{2}\,.
\end{align}
\end{subequations}
Herein, all dependence on the renormalisation scale $\zeta$ is
suppressed.  Our renormalisation procedure is described, for example,
in Ref.~\cite{Chang:2008ec}; it is straightforward to implement and
preserves the relevant symmetries.  In Eq.~\eqref{EqGap}, $l=k-q$ and
$m$ denotes the quark current mass.  The matrices
$\{\lambda^a/2\,|\,a=1,\ldots,8\}$ are the generators of SU$(3)$ colour
in the fundamental representation, $\mathcal{G}_{\mu\nu}$ denotes the
effective interaction, and $\Gamma_\nu$ is the dressed gluon--quark
vertex.  The solution of Eq.~\eqref{EqGap} can be expressed as
\begin{equation}
S(k)=\frac{1}{i\gamma\cdot k\,A(k^2)+B(k^2)}.
\label{eq:quark_prop_AB}
\end{equation}

For a meson with total momentum $P$, the Bethe--Salpeter amplitude
$\Gamma(k,P)$ satisfies~\cite{Maris:1997tm,Maris:2005tt,Krassnigg:2008bob}
\begin{equation}
\Gamma(k,P)
=
\int^\Lambda \frac{\dd^4q}{(2\pi)^4}\,
K_{q\bar q}(k,q;P)\,
S(q^+)\,
\Gamma(q,P)\,
S(q^-),
\label{eq:BSE_general}
\end{equation}
where
\begin{equation}
q^\pm=q\pm\frac{P}{2},
\end{equation}
and $K_{q\bar q}(k,q;P)$ denotes the quark--antiquark
Bethe--Salpeter kernel.

% \begin{figure}
% \centering
% \includegraphics[scale=0.8]{images/faddevv_eq.pdf}
% \caption{3-body Faddeev Equation in rainbow-ladder truncation. $\Psi$ represents the Faddeev amplitude, consisting of 128 components. Solid line, $S$, is dressed-quark propagator. The summation symbol represents the diagrams of three permutations.}
% \end{figure}
\subsection{Faddeev equation}
\label{sec:faddeev}
For a three-quark baryon, the homogeneous bound-state equation can be
written as a sum of three Faddeev components, corresponding to the three
possible choices of the spectator quark.  Suppressing the colour and
flavour indices and displaying explicitly only the Dirac--Lorentz
structure, one has
\begin{equation}
\psi^\mu_{\alpha\beta\gamma{\cal I}}(p,q,P)
=
\sum_{i=1}^{3}
\psi^{(i),\mu}_{\alpha\beta\gamma{\cal I}}(p,q,P),
\label{eq:faveq}
\end{equation}
where
\begin{subequations}
\label{eq:faveq3}
\begin{align}
\psi^{(1),\mu}_{\alpha\beta\gamma{\cal I}}(p,q,P)
&=
\int_{dk}
\left[K_{qq}(k)\right]^{\gamma^\prime\beta^\prime}_{\beta\gamma}
S_{\beta^\prime\beta^{\prime\prime}}(k_2)
S_{\gamma^\prime\gamma^{\prime\prime}}(\tilde{k}_3)
\nonumber\\
&\quad\times
\psi^\mu_{\alpha\beta^{\prime\prime}\gamma^{\prime\prime}{\cal I}}
(p^{(1)},q^{(1)},P),
\\
\psi^{(2),\mu}_{\alpha\beta\gamma{\cal I}}(p,q,P)
&=
\int_{dk}
\left[K_{qq}(k)\right]^{\alpha^\prime\gamma^\prime}_{\gamma\alpha}
S_{\gamma^\prime\gamma^{\prime\prime}}(k_3)
S_{\alpha^\prime\alpha^{\prime\prime}}(\tilde{k}_1)
\nonumber\\
&\quad\times
\psi^\mu_{\alpha^{\prime\prime}\beta\gamma^{\prime\prime}{\cal I}}
(p^{(2)},q^{(2)},P),
\\
\psi^{(3),\mu}_{\alpha\beta\gamma{\cal I}}(p,q,P)
&=
\int_{dk}
\left[K_{qq}(k)\right]^{\beta^\prime\alpha^\prime}_{\alpha\beta}
S_{\alpha^\prime\alpha^{\prime\prime}}(k_1)
S_{\beta^\prime\beta^{\prime\prime}}(\tilde{k}_2)
\nonumber\\
&\quad\times
\psi^\mu_{\alpha^{\prime\prime}\beta^{\prime\prime}\gamma{\cal I}}
(p^{(3)},q^{(3)},P).
\label{eq:faveq33}
\end{align}
\end{subequations}
Here, $\int_{dk}$ denotes a translationally invariant regularisation of
the four-dimensional momentum integral, and $K_{qq}$ is the two-quark
interaction kernel entering the three-body Faddeev equation.  The
indices $\alpha$, $\beta$, and $\gamma$ denote the Dirac indices of the
three quarks, while ${\cal I}$ is the Dirac index of the baryon.  The
additional Lorentz index $\mu$ is associated with the spin-$3/2$
Rarita--Schwinger structure. 

The three quark momenta $p_1$, $p_2$, and $p_3$ are expressed in terms
of the total momentum $P$ and two independent relative momenta $p$ and
$q$.  We choose
\begin{equation}
\begin{array}{ll}
p=(1-\eta)p_3-\eta(p_1+p_2),
&
p_1=-q-\dfrac{p}{2}+\dfrac{1-\eta}{2}P,
\\[2mm]
q=\dfrac{p_2-p_1}{2},
&
p_2=q-\dfrac{p}{2}+\dfrac{1-\eta}{2}P,
\\[2mm]
P=p_1+p_2+p_3,
&
p_3=p+\eta P,
\end{array}
\label{eq:faddeev_momenta}
\end{equation}
where $\eta$ is the momentum-partitioning parameter.  For the
equal-mass systems considered herein, $\Omega_{ccc}^{++}$ and
$\Omega_{bbb}^{-}$, we employ the symmetric choice $\eta=\frac{1}{3}$. This provides a covariant decomposition of the three-body
kinematics into the total and relative momenta.

For each of the three contributions in Eq.~\eqref{eq:faveq3}, the
interaction transfers momentum $k$ between the two participating
quarks, while the third quark acts as a spectator.  The corresponding
internal quark momenta are
\begin{equation}
k_i=p_i-k,
\qquad
\tilde{k}_i=p_i+k,
\end{equation}
and the relative momenta entering the Faddeev amplitudes are
\begin{subequations}
\label{eq:remom3}
\begin{align}
p^{(1)} &= p+k,
&
q^{(1)} &= q-\frac{k}{2},
\\
p^{(2)} &= p-k,
&
q^{(2)} &= q-\frac{k}{2},
\\
p^{(3)} &= p,
&
q^{(3)} &= q+k.
\end{align}
\end{subequations}

The Poincar\'e-covariant Faddeev amplitude of a spin-$3/2$
baryon can be expanded in a complete set of 128 Dirac--Lorentz
covariants constructed in Ref.~\cite{Sanchis-Alepuz:2011egq};
see Appendix~A and Tables~II and III therein. These Dirac--Lorentz covariants are chosen to satisfy the orthonormality
condition
\begin{equation}
\label{eq:orth}
\frac{1}{8}
\operatorname{Tr}_{D}
\left[
 \overline{\xi}^{\,\mu}_{i,\beta\alpha{\cal I}\gamma}(p,q,P)\,
 \xi^{\,\mu}_{j,\alpha\beta\gamma{\cal I}}(p,q,P)
\right]
=
\delta_{ij}.
\end{equation}
The Faddeev amplitude can therefore be expanded as
\begin{align}
\label{eq:expand}
\psi^\mu_{\alpha\beta\gamma{\cal I}}(p,q,P)
=
\sum_{l=1}^{128}
f_l(\mathfrak{T})\,
\xi^\mu_{l,\alpha\beta\gamma{\cal I}}(p,q,P),
\end{align}
where $\xi_l$ denotes the $l$-th Dirac--Lorentz covariant and
$f_l$ is its associated scalar dressing function.  On the baryon mass shell, each scalar dressing function depends on five
independent Lorentz-invariant variables.  For notational convenience,
we collect them into
$\mathfrak{T}:=(p^2,q^2,z_1,z_2,z_0)$,
where
$z_1=\widehat p\cdot\widehat P$,
$z_2=\widehat q\cdot\widehat P$, and
$z_0=\widehat p_T\cdot\widehat q_T$.
Here, $\widehat p=p/\sqrt{p^2}$ and $\widehat p^{\,2}=1$. The components of
the relative momenta transverse to the total momentum are defined by
\begin{equation}
\label{eq:transverse-momenta}
k_T^\mu
=
k^\mu
-
P^\mu\frac{P\cdot k}{P^2},
\qquad
k=p,\,q,
\end{equation}
with
$\widehat k_T=k_T/\sqrt{k_T^2}$.

In our solutions, the dominant contribution to the Faddeev amplitude is
associated with the Dirac--Lorentz covariant
$\big[S_{+}\big]^{\mu}$, defined by
\begin{align}
\label{eq:dominant_covariants}
\big[S_{+}\big]^{\mu}_{\alpha\beta\gamma{\cal I}}
:=
(\Lambda_+(\widehat P)\gamma_T^{\nu} C)_{\alpha\beta}
\otimes
\big[\mathbb{P}_{+}^{\mu\nu}(\widehat P)\big]_{\gamma{\cal I}} ,
\end{align}
%
% Hence, Eq.~\eqref{eq:dominant_covariants} represents two independent
% basis elements, corresponding to the two choices
% $\sigma=+$ and $\sigma=-$. Here,
% %
% \begin{equation}
% \Omega^\sigma(\widehat P)
% =
% \Lambda^\sigma(\widehat P)\gamma_5 C
% \otimes
% \Lambda_+(\widehat P),
% \end{equation}
%
where $C$ is the charge-conjugation matrix and
$\Lambda_\pm(\widehat P)
=
\frac{1}{2}
\left(
 \mathds{1}\pm\gamma\cdot\widehat P
\right)$ denotes the positive- or negative-energy projector. The Rarita--Schwinger projector appearing in
Eq.~\eqref{eq:dominant_covariants} is defined by
\begin{align}
\mathbb{P}_\pm^{\mu\nu}(P)
&=
\Lambda_\pm(\widehat P)
\left(
 T_P^{\mu\nu}
 -\frac{1}{3}
 \gamma_T^\mu\gamma_T^\nu
\right),
\label{eq:RS_projector}
\end{align}
with
\begin{equation}
T_P^{\mu\nu}
=
\delta^{\mu\nu}
-\widehat P^\mu\widehat P^\nu,
\qquad
\gamma_T^\mu
=
T_P^{\mu\nu}\gamma^\nu .
\label{eq:transverse_objects}
\end{equation}
The hat denotes a normalised four-vector. The projector
$\mathbb{P}^{\mu\nu}$ selects the positive-energy,
spin-$3/2$ component and satisfies the transversality conditions
\begin{equation}
\widehat P^\mu\mathbb{P}_\pm^{\mu\nu}(P)=0,
\qquad
\gamma^\mu\mathbb{P}_\pm^{\mu\nu}(P)=0.
\end{equation}

\begin{figure*}[!t]
	\centering
	% ==================== 第一行 ====================
	\begin{minipage}[t]{0.49\textwidth}
		\centering

		\makebox[\linewidth][l]{%
			\textsf{\large A}%
			\hfill
			\textsf{\large$\Omega_{ccc}^{++}$}%
			\hfill
			\phantom{\textsf{\large A}}%
		}\\[-0.4em]

		\begin{minipage}[c]{1.8em}
			\centering
			\rotatebox{90}{%
				\textsf{$c^{0,0,0}_1\bigl(p^2,q^2\bigr)$}%
			}
		\end{minipage}
		\begin{minipage}[c]{0.85\textwidth}
			\centering
			\includegraphics[width=\linewidth]
			{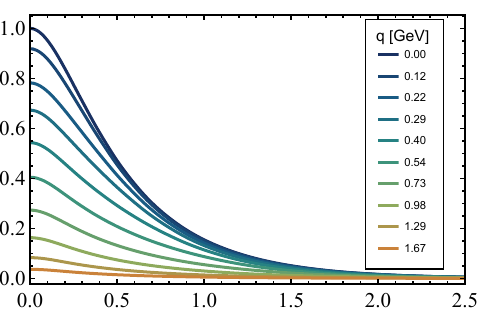}\\[-0.5em]
			\textsf{$p/\mathrm{GeV}$}
		\end{minipage}

	\end{minipage}
	\hfill
	\begin{minipage}[t]{0.49\textwidth}
		\centering

		\makebox[\linewidth][l]{%
			\textsf{\large B}%
			\hfill
			\textsf{\large$\Omega_{bbb}^{-}$}%
			\hfill
			\phantom{\textsf{\large B}}%
		}\\[-0.4em]

		\begin{minipage}[c]{1.8em}
			\centering
			\rotatebox{90}{%
				\textsf{$c^{0,0,0}_1\bigl(p^2,q^2\bigr)$}%
			}
		\end{minipage}
		\begin{minipage}[c]{0.85\textwidth}
			\centering
			\includegraphics[width=\linewidth]
			{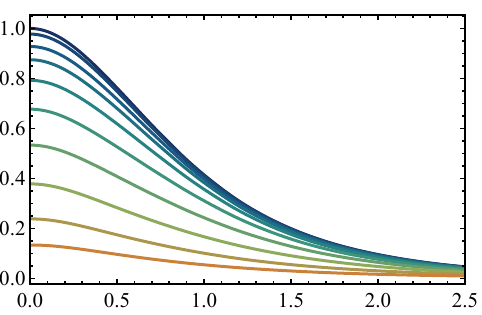}\\[-0.5em]
			\textsf{$p/\mathrm{GeV}$}
		\end{minipage}

	\end{minipage}

	\vspace{1.0em}

	\begin{minipage}[t]{0.49\textwidth}
		\centering

		\makebox[\linewidth][l]{%
			\textsf{\large C}%
			\hfill
			%\textsf{ $\Omega_{ccc}^{++}$}%
			\hfill
			\phantom{\textsf{\large C}}%
		}\\[-0.4em]

		\begin{minipage}[c]{1.8em}
			\centering
			\rotatebox{90}{%
				\textsf{$c^{n_1,0,0}_1\bigl(p^2,q^2=0\bigr)$}%
			}
		\end{minipage}
		\begin{minipage}[c]{0.85\textwidth}
			\centering
			\includegraphics[width=\linewidth]
			{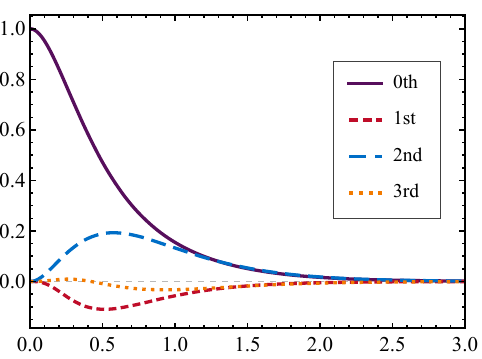}\\[-0.5em]
			\textsf{$p/\mathrm{GeV}$}
		\end{minipage}

	\end{minipage}
	\hfill
	\begin{minipage}[t]{0.49\textwidth}
		\centering

		\makebox[\linewidth][l]{%
			\textsf{\large D}%
			\hfill
			%\textsf{$c_l\bigl(p^2,q^2=0,n_1,0,0\bigr)$}%
			\hfill
			\phantom{\textsf{\large D}}%
		}\\[-0.4em]

		\begin{minipage}[c]{1.8em}
			\centering
			\rotatebox{90}{%
				\textsf{$c^{n_1,0,0}_1\bigl(p^2,q^2=0\bigr)$}%
			}
		\end{minipage}
		\begin{minipage}[c]{0.85\textwidth}
			\centering
			\includegraphics[width=\linewidth]
			{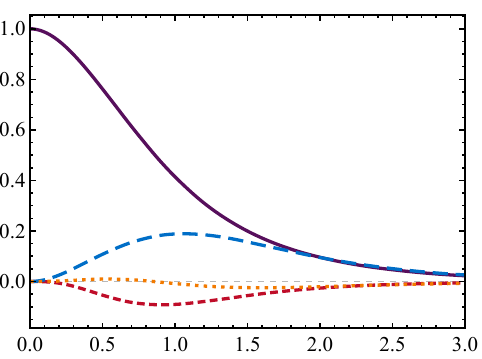}\\[-0.5em]
			\textsf{$p/\mathrm{GeV}$}
		\end{minipage}

	\end{minipage}
\caption{Panels A and B show the zeroth Chebyshev moments of the coefficient
functions $c_i$ associated with the dominant Faddeev amplitude, plotted
as functions of the relative momenta $p$ and $q$ for the
$\Omega_{ccc}^{++}$ and $\Omega_{bbb}^{-}$ baryons, respectively.
The amplitudes are normalized such that
$c^{0,0,0}_1\bigl(p^2=0,q^2=0\bigr)=1$. Panels C and D display the first four Chebyshev moments with respect to
the angular variable $z_1$ of the dominant amplitude for the
$\Omega_{ccc}^{++}$ and $\Omega_{bbb}^{-}$ baryons, respectively.
}
	\label{fig:ccc-bbb-amplitudes}
\end{figure*}
To facilitate the numerical solution of the Faddeev equation and the
subsequent evaluation of matrix elements, the dependence of each
coefficient function on the three angular variables is expanded in a
tensor-product basis of Chebyshev polynomials. Each coefficient function is reconstructed according to
\begin{align}
f_l\bigl(\mathfrak{T}\bigr)
=&
\sum_{n_1=0}^{N_1-1}
\sum_{n_2=0}^{N_2-1}
\sum_{n_0=0}^{N_0-1}
T_{n_1}(z_1)\,
T_{n_2}(z_2)\,
T_{n_0}(z_0)
\nonumber\\
&\times
c^{n_1,n_2,n_0}_l\bigl(p^2,q^2\bigr),
\label{eq:chebyshev-reconstruction}
\end{align}
where $N_1$, $N_2$, and $N_0$ denote the truncation orders associated
with the angular variables $z_1$, $z_2$, and $z_0$, respectively.
The coefficients
$c^{n_1,n_2,n_0}_l\bigl(p^2,q^2\bigr)$ are the corresponding Chebyshev moments
of the coefficient function associated with the covariant labelled by
$l$.

In the Euclidean convention employed here, the normalized Chebyshev
basis is defined by
\begin{align}
T_n(z)
&=
\mathrm{i}^{\,n}\,\frac{n}{2}
\sum_{k=0}^{\lfloor n/2\rfloor}
(-1)^k
\frac{(n-k-1)!}{(n-2k)!\,k!}
(2z)^{n-2k},
\ n\geq 1.
\label{eq:chebyshev-definition}
\\
T_0(z)
&=
\frac{1}{\sqrt{2}}.
\end{align}
where $\mathrm{i}$ denotes the imaginary unit.
The factors $\mathrm{i}^n$ arise from the Euclidean angular convention
adopted in the parametrization of the relative momenta. In practice, the rapid
suppression of the higher Chebyshev moments permits the angular
dependence of the Faddeev amplitude to be represented accurately with
a comparatively small number of terms.

\begin{table}[t]
\centering
\caption{
Calculated masses and leptonic decay constants of selected heavy
mesons, together with the masses of the ground-state triply heavy
baryons. Available experimental and lQCD results are included
for comparison. For the $\Omega_{ccc}^{++}$, we quote both the most
precise lattice determination currently available and the result
obtained in the lattice study of its electromagnetic form factors.
All dimensionful quantities are given in GeV.
}
\label{TabHeavySystems}

\renewcommand{\arraystretch}{1.10}
\setlength{\tabcolsep}{5.0pt}

% Meson sector
\begin{tabular}{lcccc}
\hline\hline

Meson
& $m_{\rm calc.}$
& $m_{\rm expt.}$
& $f_{\rm calc.}$
& $f_{\rm expt.}$
\\
\midrule
$\eta_c$
& 2.98
& 2.98
& 0.28
& 0.24
\\
$J/\psi$
& 3.12
& 3.10
& 0.30
& 0.29
\\
$\eta_b$
& 9.40
& 9.40
& 0.57
& --
\\
$\Upsilon$
& 9.50
& 9.46
& 0.54
& 0.51
\\
\bottomrule
\end{tabular}

\vspace{0.8em}

% Baryon sector

\begin{tabular}{lcc}
\toprule
\midrule
Baryon
& $m_{\rm calc.}$
& $m_{\rm lQCD}$
\\
\midrule
$\Omega_{ccc}^{++}$
& 4.76
& 4.793(9)~\cite{Dhindsa:2024ccc}
\\
&
& 4.769(6)~\cite{Can:2015exa}\\
$\Omega_{bbb}^{-}$
& 14.35 % Please verify: the original draft contained 9.40.
& 14.371(12)~\cite{Meinel:2010bbb}
\\
\bottomrule
\end{tabular}
\end{table}

\subsection{Interaction kernel and rainbow--ladder truncation}

Solving the Bethe--Salpeter and Faddeev equations requires the dressed-quark
propagator and the interaction kernels between the constituent quarks. In
this work, these quantities are obtained using the rainbow--ladder (RL)
truncation. It is the leading-order term in a systematic
symmetry-preserving truncation scheme and provides the most widely used and
economical framework for the simultaneous study of mesons and baryons.
When the rainbow approximation in the quark gap equation is combined
consistently with the ladder approximation in the bound-state equations,
the relevant vector and axial-vector Ward--Green--Takahashi identities are
preserved.

In the rainbow approximation, the product of the dressed-gluon propagator
and dressed gluon--quark vertex is replaced according to
\begin{equation}
\label{eq:rainbow}
g^{2}D_{\mu\nu}(k)\Gamma_{\nu}(q,p)
\longrightarrow
D_{\mu\nu}^{\mathrm{free}}(k)\,
k^{2}\mathcal{G}(k^{2})\gamma_{\nu}.
\end{equation}
Consistency with the rainbow approximation requires the interaction kernels
in the meson Bethe--Salpeter and baryon Faddeev equations to be treated in
the corresponding ladder approximation. For either the $q\bar q$ or $qq$ channel, the resulting two-body
kernel can be written uniformly as
\begin{equation}
\label{eq:ladder-kernel}
K_{\mathcal{H}}^{\mathrm{RL}}(k)
=
- \mathbb{C}_{\mathcal{H}} Z_{2}^{2}\,
k^{2}\mathcal{G}(k^{2})
D_{\mu\nu}^{\mathrm{free}}(k)\,
\gamma_{\mu}\otimes\gamma_{\nu},
\end{equation}
where \(\mathbb{C}_{q\bar q}=4/3\) is obtained by projecting the
quark--antiquark interaction onto the colour-singlet channel, whereas
\(\mathbb{C}_{qq}=2/3\) follows from projection of the quark--quark interaction onto
the attractive colour-\(\bar{\mathbf 3}\) channel. Thus, the strength of the
pairwise interaction in the colour-antitriplet \(qq\) channel is one half of
that in the colour-singlet \(q\bar q\) channel.

More generally, the RL truncation retains only the leading
$\gamma_\mu$ structure of the dressed quark--gluon vertex, with its
dominant dressing effects absorbed into the effective interaction
$\mathcal{G}(k^2)$.  It therefore omits the remaining vertex
covariants, including anomalous chromomagnetic-moment (ACM) structures~\cite{Chang:2010jqj,Binosi:2016rxz,Qin:2020rad,Xu:2022kng},
as well as non-ladder interaction topologies. Such ACM
structures can in turn generate a sizeable anomalous magnetic moment
(AMM) of the dressed quark, which can produce visible corrections to
hadron electromagnetic form factors, particularly in the magnetic and
higher-multipole sectors~\cite{Xing:2021dwe,Wilson:2011aa,Albino:2026koj}. These limitations, however, are
expected to be less severe for the systems studied herein.
Calculations employing more complete quark--gluon vertices indicate
that low-lying ground states are comparatively insensitive to the
detailed tensor structure of the interaction, whereas excited states
are generally more sensitive~\cite{SanchisAlepuz:2015tha,Fischer:2014xha}.
Moreover, corrections beyond RL decrease with increasing quark mass
and become strongly suppressed in the heavy-quark limit~\cite{Qin:2019oar,Xu:2024fun,Hagel:2025ngi}.  Hence, for the ground-state
$\Omega_{ccc}$ and $\Omega_{bbb}$ baryons, the omitted vertex
structures, including the ACM term, are expected to provide
subleading corrections.  
% Nevertheless, their possible influence on
% spin-sensitive observables and higher multipole form factors
% constitutes a residual truncation uncertainty.

Hence, the meson and baryon bound-state equations
are constructed from the same symmetry-consistent effective interaction~\cite{Qin:2011dd}, $viz$.,
\begin{subequations}
	\begin{align}
		\mathcal{G}(s)&=\frac{8 \pi^{2}}{\omega^{4}}D e^{-s/ \omega^{2}}+ \frac{8 \pi^{2} \gamma_{m} \mathcal{F}(s)}{\ln[\tau+(1+s/ \Lambda_{\mathrm{Q C D}}^{2})^{2}]},\label{E4}\\
		\mathcal{F}(s)&=[1-e^{-s /(4 m_{t}^{2})}]/ s,
		\label{E5}
	\end{align}
\end{subequations}
 with $m_{t}=0.5\,\mathrm{GeV}, \tau=e^{2}-1, \Lambda_{\mathrm{QCD}}=0.36\,\mathrm{GeV}, \gamma_{m}=4/\beta_0$, $\beta_0=11-(2/3)n_f$, $n_f=5$.
The infrared parameters controlling the strength and range of the
interaction are chosen as
\begin{equation}
\label{eq:interaction_parameters}
\omega=0.8\,\mathrm{GeV},
\qquad
D=0.27\,\mathrm{GeV}^{2}.
\end{equation}
We employ a mass-independent renormalization scheme with the
renormalization point \(\zeta=\zeta_{19}=19\,\mathrm{GeV}\) and the current-quark mass are $m_c^{\zeta_{19}}=0.82\,\mathrm{GeV}$ and $m_b^{\zeta_{19}}=3.65\,\mathrm{GeV}$.

\subsection{Numerical solution of BSE and Faddeev equation}
With the ingredients specified above, the BSE,
Eq.~\eqref{eq:BSE_general}, and the three-body Faddeev equation,
Eq.~\eqref{eq:faveq}, can be solved numerically.  Despite their
different explicit structures, both are homogeneous bound-state
equations whose solutions occur at discrete timelike values of the
total momentum corresponding to hadron masses.

\begin{figure*}
\centering
\includegraphics[scale=0.7]{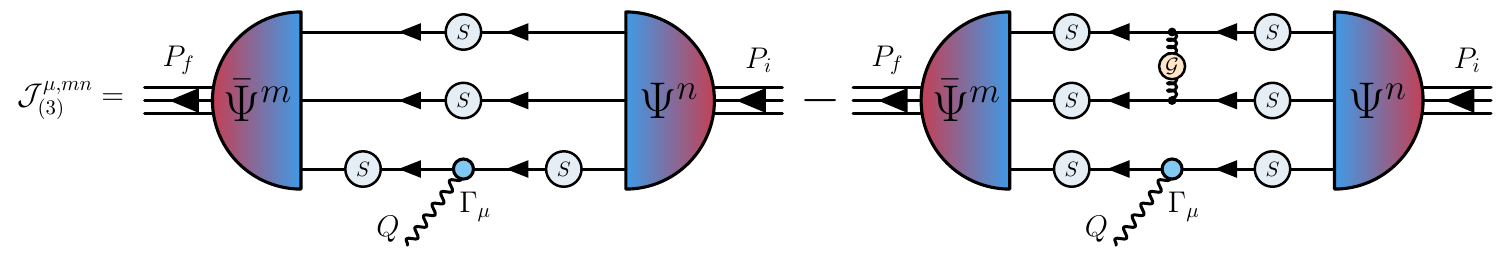}
\caption{\( \Gamma_\mu \) represents the dressed quark-photon vertex. The subscript
``$(3)$'' labels the contribution in which the
photon couples to the third quark leg. The superscripts $m$ and $n$ denote the outgoing and incoming Rarita--Schwinger
indices of the spin-$3/2$ baryon. This diagram illustrates the interaction between the photon and one of the quarks in the triply-heavy baryon, and therefore, there are three such diagrams in total.}
\label{fig:three_body_current}
\end{figure*}

Numerical solutions of meson Bethe--Salpeter equations are by now well
established and have found a wide range of applications~\cite{Xiao:2025cqz,Chang:2013pq,Chen:2012qr,Chang:2013nia}.
By contrast, solving the three-body Faddeev equation remains
considerably more demanding because of the larger number of momentum
variables and Dirac--Lorentz covariants involved.  In the expansion of
Eq.~\eqref{eq:expand}, we employ 40 grid points for each of the radial
variables $\{p^2,q^2\}$ and 6 points for each of the angular variables
$\{z_1,z_2,z_0\}$.  For the momentum integrations entering
Eq.~\eqref{eq:faveq3}, we use 40 quadrature points for the radial
integration and 12 points for each angular integration.  This
discretisation provides stable and well-converged baryon masses.  The
comparatively dense momentum grids are employed primarily to improve
the numerical precision required for the subsequent calculation of
electromagnetic form factors.

As a first assessment of the interaction and heavy-quark inputs, the
masses and leptonic decay constants of the relevant heavy mesons,
together with the calculated masses of the ground-state triply-heavy
baryons, are collected in Table~\ref{TabHeavySystems}.  For the
$\Omega_{ccc}^{++}$, we quote both the most precise lattice
determination currently available~\cite{Dhindsa:2024ccc} and the mass
reported in Ref.~\cite{Can:2015exa}, whose lattice-QCD (lQCD) results for the
electromagnetic form factors will be used for comparison below.  For
the $\Omega_{bbb}^{-}$, our result is compared with the lQCD
prediction of Ref.~\cite{Meinel:2010bbb}.
Although the rainbow--ladder truncation omits the ACM and other
nonleading tensor structures of the fully dressed quark--gluon vertex,
together with the corresponding corrections to the bound-state kernels,
Table~\ref{TabHeavySystems} shows that it nevertheless provides a good
description of the masses of ground-state pseudoscalar and vector heavy
mesons, as well as those of ground-state triply-heavy baryons with
$J^P=3/2^+$.  In the present framework, the dressed quark--photon
vertex is obtained from an inhomogeneous BSE in the $J^{PC}=1^{--}$
channel and consequently develops poles at the corresponding
vector-meson masses.  These poles provide a dynamical realisation of
vector-meson dominance and play an important role in determining the
momentum dependence of hadron electromagnetic form factors.  Hence, a
realistic description of the vector-meson spectrum, together with the
symmetry-consistent treatment of the quark--photon vertex and the
bound-state amplitudes, provides an important consistency check for the
calculation of the triply-heavy baryon form factors presented below.

Figure~\ref{fig:ccc-bbb-amplitudes} illustrates the coefficient function
associated with the dominant Faddeev covariant,
Eq.~\eqref{eq:dominant_covariants}, for the
$\Omega_{ccc}^{++}$ and $\Omega_{bbb}^{-}$ baryons.
Panels~A and B display its zeroth Chebyshev moment as a function of the
relative momentum $p$ for a representative set of values of $q$.
At fixed $q$, the coefficient functions decrease monotonically with
increasing $p$, while increasing $q$ suppresses their overall magnitude.
A comparison of the two panels shows that the corresponding
$\Omega_{bbb}^{-}$ component decreases more slowly with increasing
relative momentum and retains appreciable strength over a broader
momentum range than its $\Omega_{ccc}^{++}$ counterpart.  The dominant
Faddeev component of the $\Omega_{bbb}^{-}$ is therefore harder in
momentum space.

Panels C and D in Fig.~\ref{fig:ccc-bbb-amplitudes} show the first four
Chebyshev moments with respect to the angular variable $z_1$.  In both
systems, the zeroth moment provides the dominant contribution, while the
nonzero moments are substantially smaller.  The first and second moments
describe the leading angular dependence and appear with opposite signs over
most of the displayed momentum range, whereas the third moment is already
strongly suppressed.
% {\color{red} Although the moments do not decrease strictly monotonically with their
% order, their overall hierarchy demonstrates that the angular dependence
% is smooth and can be accurately represented using only the first few
% Chebyshev moments.}

\section{Electromagnetic Form Factors of spin-$3/2$ baryons}
\label{sec:spin32_EFFs}
The microscopic electromagnetic current is obtained by gauging the
three-quark scattering matrix and identifying its residue at the
incoming and outgoing baryon poles.  In this construction, the photon
couples to all possible internal propagators and interaction kernels,
yielding a conserved hadronic current provided that the relevant
vector Ward--Takahashi identities are preserved~\cite{Eichmann:2011ec,Eichmann:2026correlation,SanchisAlepuz:2013}.

To specify the elastic kinematics, we introduce the average baryon
momentum $P$ and photon momentum $Q$ through
\begin{equation}
P=\frac{P_f+P_i}{2},
\qquad
Q=P_f-P_i,
\qquad
P_i^2=P_f^2=-M^2,
\end{equation}
where $P_i$ and $P_f$ are the incoming and outgoing baryon momenta,
respectively.  The on-shell conditions imply $P\cdot Q=0$.

The Lorentz structure of the corresponding on-shell current is
independent of its particular microscopic realization.  For an
on-shell spin-$3/2$ baryon, the elastic electromagnetic current is
completely specified by four independent covariant form factors,
$F_i(Q^2)$, $i=1,\ldots,4$~\cite{Nozawa:1990,Nicmorus:2010,SanchisAlepuz:2013}.  In the Euclidean
conventions employed here, it can be written as
\begin{align}
\mathcal{J}^{\mu,mn}
&=
\mathbb{P}^{mm'}(P_f)
\Bigg\{
\left(
 [F_1+F_2]\,i\gamma^\mu
 -F_2\frac{P^\mu}{M}
\right)\delta^{m'n'}
\nonumber\\
&
+
\left(
 [F_3+F_4]\,i\gamma^\mu
 -F_4\frac{P^\mu}{M}
\right)
\frac{Q^{m'}Q^{n'}}{4M^2}
\Bigg\}
\mathbb{P}^{n'n}(P_i),
\label{eq:spin32_current}
\end{align}
The four covariant form factors $F_i(Q^2)$ can be recast in terms of
the electric-monopole, magnetic-dipole, electric-quadrupole, and
magnetic-octupole form factors,
$G_{E0}$, $G_{M1}$, $G_{E2}$, and $G_{M3}$, respectively.

Within the present truncation, one representative contribution to the
microscopic three-quark current is
\begin{align}
\label{ImFF}
\left[\mathcal{J}_{(3)}^{\mu,mn}\right]_{{\cal I}'{\cal I}}
&=
\iint_{pq}
\bar{\psi}^{m}_{\beta'\alpha'{\cal I}'\gamma'}
(p_f,q_f,P_f)\nonumber\\
&\times
S_{\alpha'\alpha}(p_1)
S_{\beta'\beta}(p_2)
\chi_{\gamma'\gamma}^\mu(p_3,Q)\nonumber\\
&\times
\left[
 \psi^{n}_{\alpha\beta\gamma{\cal I}}(p_i,q_i,P_i)
 -
 \psi^{n,(3)}_{\alpha\beta\gamma{\cal I}}(p_i,q_i,P_i)
\right].
\end{align}
Here, $\psi^{n,(3)}_{\alpha\beta\gamma{\cal I}}$ denotes the third
Faddeev component defined in Eq.~\eqref{eq:faveq33}. The variables
$p_i,q_i$ and $p_f,q_f$ denote the incoming and outgoing relative
momenta, respectively.  Furthermore,
\begin{equation}
\chi^\mu(p_3,Q)
=
S(p_3^+)\,
\Gamma^\mu(p_3,Q)\,
S(p_3^-),
\qquad
p_3^\pm=p_3\pm\frac{Q}{2},
\end{equation}
is the dressed quark--photon wave function, with $\Gamma^\mu$ obtained
from the corresponding inhomogeneous BSE.  The spectator-quark
momenta are
\begin{align}
p_1&=-q-\frac{p}{2}+\frac{P}{3},
&
p_2&=q-\frac{p}{2}+\frac{P}{3}.
\end{align}

\begin{figure}[t!]
\centering
\begin{minipage}[t]{0.49\textwidth}
	\centering

	\makebox[\linewidth][l]{%
		\textsf{\large A:\ $\Omega_{ccc}^{++}$}%
		\hfill
		\phantom{\textsf{\large A:\ $\Omega_{ccc}^{++}$}}%
	}\\[-0.4em]

	\begin{minipage}[c]{1.8em}
		\centering
		\rotatebox{90}{%
			\textsf{$G_{E0}(Q^2)/G_{E0}(0)$}%
		}
	\end{minipage}
	\begin{minipage}[c]{0.85\textwidth}
		\centering
		\includegraphics[width=\linewidth]
		{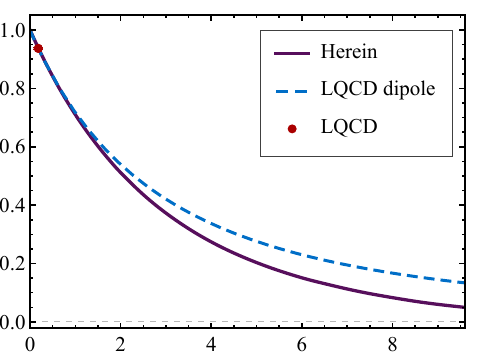}\\[-0.5em]
		\textsf{$Q^2/\mathrm{GeV}^2$}
	\end{minipage}

\end{minipage}
\hfill
% ============================================================
% Panel B: Omega_bbb
% ============================================================
\begin{minipage}[t]{0.49\textwidth}
	\centering

	\makebox[\linewidth][l]{%
			\textsf{\large B:\ $\Omega_{bbb}^{-}$}%
		\hfill
		\phantom{\textsf{\large B}}%
	}\\[-0.4em]

	\begin{minipage}[c]{1.8em}
		\centering
		\rotatebox{90}{%
			\textsf{$G_{E0}(Q^2)/G_{E0}(0)$}%
		}
	\end{minipage}
	\begin{minipage}[c]{0.85\textwidth}
		\centering
		\includegraphics[width=\linewidth]
		{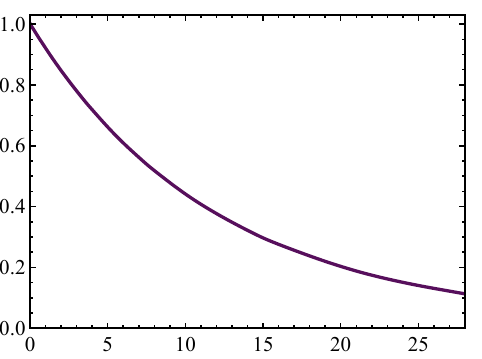}\\[-0.5em]
		\textsf{$Q^2/\mathrm{GeV}^2$}
	\end{minipage}
\end{minipage}
\caption{Normalized electric-monopole form factors, $G_{E0}(Q^2)/G_{E0}(0)$, of (\textsf{A})~$\Omega_{ccc}^{++}$ and (\textsf{B})~$\Omega_{bbb}^{-}$. The purple solid curves show our results. For $\Omega_{ccc}^{++}$, the red point denotes the available lQCD result at $Q^2=0.183~\mathrm{GeV}^2$. The corresponding dashed curve is the dipole reconstruction obtained from this lQCD point.}
\label{fig:GE0-ccc-bbb}
\end{figure}
As illustrated in Fig.~\ref{fig:three_body_current}, the subscript in $\mathcal{J}_{(3)}^{\mu,mn}$ labels the contribution in which the
photon couples to the third quark leg. Two analogous contributions are
obtained by coupling the photon to the first and second quark legs.
The complete three-quark current is therefore
\begin{equation}
\mathcal{J}^{\mu,mn}
=
\sum_{a=1}^{3}
q_a\,
\mathcal{J}_{(a)}^{\mu,mn},
\label{eq:total_three_quark_current}
\end{equation}
where $q_a$ denotes the electric charge of the $a$-th
quark in units of the elementary charge $e$. For a baryon composed of three identical quarks and with a symmetric
momentum partition, the three terms are related by permutations and
give identical contributions apart from their electric-charge
factors.

The four covariant form factors can be recast in terms of the physical
electromagnetic multipole form factors: the electric monopole,
$G_{E0}$, magnetic dipole, $G_{M1}$, electric quadrupole,
$G_{E2}$, and magnetic octupole, $G_{M3}$.  They are related to the
covariant form factors by~\cite{Nozawa:1990,SanchisAlepuz:2013}
\begin{subequations}
\begin{align}
G_{E0}(Q^2)
=&
\left(1+\frac{2\tau}{3}\right)
\left(F_1-\tau F_2\right)
-\frac{\tau}{3}(1+\tau)
\left(F_3-\tau F_4\right),
\label{eq:GE0_def}
\\
G_{M1}(Q^2)
=&
\left(1+\frac{4\tau}{5}\right)
\left(F_1+F_2\right)
-\frac{2\tau}{5}(1+\tau)
\left(F_3+F_4\right),
\label{eq:GM1_def}
\\
G_{E2}(Q^2)
=&
\left(F_1-\tau F_2\right)
-\frac{1}{2}(1+\tau)
\left(F_3-\tau F_4\right),
\label{eq:GE2_def}
\\
G_{M3}(Q^2)
=&
\left(F_1+F_2\right)
-\frac{1}{2}(1+\tau)
\left(F_3+F_4\right),
\label{eq:GM3_def}
	\end{align}
\label{eq:GM_def}
\end{subequations}
where $\tau=Q^2/(4M^2)$.  Here and below, the arguments $Q^2$ of the
$F_i$ on the right-hand sides are left implicit.  Nonzero values of
$G_{E2}$ and $G_{M3}$ signal nonspherical components in the electric
and magnetic distributions, respectively.

% At zero momentum transfer, the four multipole form factors define
% the baryon charge, magnetic dipole moment, electric quadrupole moment, and magnetic octupole moment:
% %
% \begin{align}
% \frac{q_B}{e}
% &=
% G_{E0}(0),
% \label{eq:spin32_charge}
% \\
% \mu_B
% &=
% \frac{e}{2M}G_{M1}(0),
% \label{eq:spin32_magnetic_moment}
% \\
% \mathcal{Q}_B
% &=
% \frac{e}{M^2}G_{E2}(0),
% \label{eq:spin32_quadrupole_moment}
% \\
% \mathcal{O}_B
% &=
% \frac{e}{2M^3}G_{M3}(0).
% \label{eq:spin32_octupole_moment}
% \end{align}
%
% Equivalently, the magnetic moment expressed in nuclear magnetons is
% %
% \begin{equation}
% \frac{\mu_B}{\mu_N}
% =
% \frac{m_N}{M}G_{M1}(0).
% \label{eq:spin32_magnetic_moment_muN}
% \end{equation}
%

\section{Results and discussion}
\label{sec:Results}
Using the Faddeev amplitudes obtained above, we evaluate the
three-quark electromagnetic current in
Eq.~\eqref{eq:total_three_quark_current}.  The projection operators
introduced in Appendix~\ref{sec:extraction} are then employed to
extract the electromagnetic multipole form factors defined in
Eq.~\eqref{eq:GM_def}.

\begin{table}[t]
\centering
\caption{
Electric and magnetic radii of the triply heavy baryons. The superscripts $c$ and $b$ denote $\Omega_{ccc}^{++}$ and $\Omega_{bbb}^{-}$, respectively. All dimensionful quantities are given in fm.}
\label{tab:radii}
\begin{tabular}{lcccc}
\hline\hline
& $r^c_{E_0}$ 
& $r^c_{M_1}$&$r^b_{E_0}$ 
& $r^b_{M_1}$ \\
\hline
Herein
&$0.287(1)$& $0.279(1)$& $0.140(2)$& $0.138(3)$ \\
lQCD
& $0.290(5)$
& $0.290(5)$& $-$
& $-$ \\
\bottomrule
\end{tabular}
\end{table}

\begin{figure}[t]
	\centering

	% ------------------------------------------------------------
	% Panel A: Omega_ccc
	% ------------------------------------------------------------
	\begin{minipage}[t]{0.49\textwidth}
		\centering

		\makebox[\linewidth][l]{%
			\textsf{\large A:\ $\Omega_{ccc}^{++}$}%
			\hfill
			\phantom{\textsf{\large A}}%
		}\\[-0.4em]

		\begin{minipage}[c]{1.8em}
			\centering
			\rotatebox{90}{%
				\textsf{$G_{M1}(Q^2)/G_{M1}(0)$}%
			}
		\end{minipage}
		\begin{minipage}[c]{0.85\textwidth}
			\centering
			\includegraphics[width=\linewidth]
			{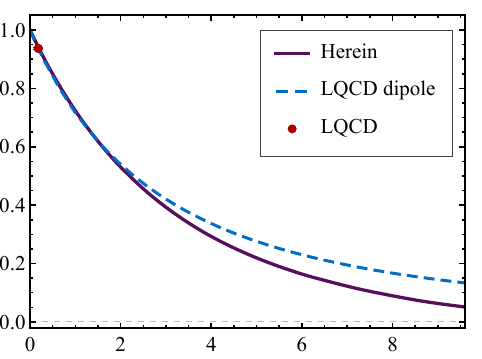}\\[-0.5em]
			\textsf{$Q^2/\mathrm{GeV}^2$}
		\end{minipage}

	\end{minipage}
	\hfill
	% ------------------------------------------------------------
	% Panel B: Omega_bbb
	% ------------------------------------------------------------
	\begin{minipage}[t]{0.49\textwidth}
		\centering

		\makebox[\linewidth][l]{%
			\textsf{\large B:\ $\Omega_{bbb}^{-}$}%
			\hfill
			\phantom{\textsf{\large B}}%
		}\\[-0.4em]

		\begin{minipage}[c]{1.8em}
			\centering
			\rotatebox{90}{%
				\textsf{$G_{M1}(Q^2)/G_{M1}(0)$}%
			}
		\end{minipage}
		\begin{minipage}[c]{0.85\textwidth}
			\centering
			\includegraphics[width=\linewidth]
			{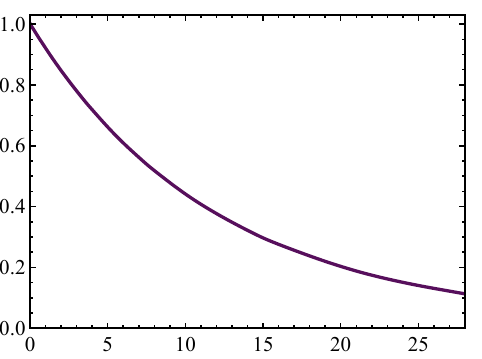}\\[-0.5em]
			\textsf{$Q^2/\mathrm{GeV}^2$}
		\end{minipage}

	\end{minipage}

	\caption{
	Normalized magnetic-dipole form factors,
	$G_{M1}(Q^2)/G_{M1}(0)$, of
	(\textsf{A})~$\Omega_{ccc}^{++}$ and
	(\textsf{B})~$\Omega_{bbb}^{-}$.
	The purple solid curves show our results.
	For $\Omega_{ccc}^{++}$, the red point denotes the available lQCD
	result at $Q^2=0.183~\mathrm{GeV}^2$.
	The corresponding dashed curve is obtained using the momentum dependence
	assumed in the lQCD analysis.
	}
	\label{fig:GM1-ccc-bbb}
\end{figure}

\subsection{Electric monopole form factor}
\label{sec:GE0}
For the triply-heavy systems considered herein, charge conservation
requires
\begin{equation}
G_{E0}^{\Omega_{ccc}^{++}}(0)=3q_c=2,
\qquad
G_{E0}^{\Omega_{bbb}^{-}}(0)=3q_b=-1,
\label{eq}
\end{equation}
where $q_c=2/3$ and $q_b=-1/3$ denote the quark charges in units of
the elementary charge. Using the Nakanishi-normalized Faddeev
amplitudes, our direct numerical calculation reproduces these charge
normalizations to within $1\%$. In the results presented below, we
therefore impose the exact charge normalization,
$G_{E0}^{\Omega_{ccc}^{++}}(0)=2$ and
$G_{E0}^{\Omega_{bbb}^{-}}(0)=-1$.
Figure~\ref{fig:GE0-ccc-bbb} displays the normalized electric monopole
form factors, $G_{E0}(Q^2)/G_{E0}(0)$, for
$\Omega_{ccc}^{++}$ and $\Omega_{bbb}^{-}$ in panels
(\textsf{A}) and (\textsf{B}), respectively.

For $\Omega_{ccc}^{++}$, the published lQCD result is available at only
one nonzero momentum transfer,
$Q^2=0.183~\mathrm{GeV}^2$~\cite{Can:2015exa}.
In Fig.~\ref{fig:GE0-ccc-bbb}(\textsf{A}), the red point denotes this
lQCD result, while the dashed curve represents a dipole reconstruction
constrained by this point together with the charge normalization at
$Q^2=0$.  The purple solid curve represents our direct calculation.
At low momentum transfer, our result agrees very well with the available
lQCD result and the corresponding dipole reconstruction.

As $Q^2$ increases, however, the two curves gradually separate, with
our result falling below the dipole reconstruction.  This growing
difference illustrates the uncertainty associated with extrapolating a
simple dipole parametrization beyond the momentum region constrained by
the lQCD calculation.  In the present case, the dipole Ansatz yields a
harder form factor and hence larger values of $G_{E0}(Q^2)$ at high
momentum transfer than obtained from the direct calculation.  A
qualitatively similar tendency was found in the unified DSE study of
the pion and kaon elastic electromagnetic form factors in
Ref.~\cite{Yao:2024drm}, where a simple single-pole vector-meson-dominance (VMD) parametrization
likewise produces a harder large-$Q^2$ behavior than the corresponding
direct calculation.

\begin{table}[t]
\centering
\caption{
Predictions for the magnetic moments of the ground-state triply heavy
baryons, in units of the nuclear magneton $\mu_N=e/(2m_p)$, where
$m_p$ is the proton mass.  Results from the nonrelativistic Faddeev
calculation using the AL1 interaction (NR Faddeev)~\cite{SilvestreBrac:1996},
the hypercentral constituent quark model (hCQM)~\cite{Patel:2008},
the effective-mass scheme (EMS) and screened-charge scheme
(SCS)~\cite{Dhir:2009}, the bag model~\cite{Bernotas:2012},
and lQCD~\cite{Can:2015exa} are included for comparison.
}
\label{tab:heavy-baryon-magnetic}
\renewcommand{\arraystretch}{1.12}
\setlength{\tabcolsep}{7pt}
\begin{tabular}{lcc}
\hline\hline
$\mu_{B}/\mu_N$
& $\Omega_{ccc}^{++}$
& $\Omega_{bbb}^{-}$
\\
\midrule

Herein
& 1.26
& $-0.207$\\

NR Faddeev (AL1)
& 1.023
& $-0.180$
\\

hCQM
& $1.176(19)$
& $-0.194(1)$
\\

EMS
& 1.10
& --
\\

SCS
& 1.16
& --
\\

Bag model
& 1.138
& $-0.194$
\\
lQCD
& 0.676(5)
& --
\\
\bottomrule
\end{tabular}
\end{table}
A comparison between panels (\textsf{A}) and (\textsf{B}) in
Fig.~\ref{fig:GE0-ccc-bbb} shows that, as the quark mass increases
from the charm to the bottom sector, the electric form factor becomes
harder, indicating a more compact spatial distribution for the
$\Omega_{bbb}^{-}$. A more
quantitative measure of this behaviour is provided by the corresponding
mean-square radius, which is determined by the slope of the form factor
at the origin:
\begin{equation}
\langle r_{E_0}^{2}\rangle
=
-\frac{6}{G_{E0}(0)}
\left.
\frac{dG_{E0}(Q^2)}{dQ^2}
\right|_{Q^2=0},
\label{eq:GE0-radius}
\end{equation}
where the corresponding root-mean-square(rms) radius is defined as
\begin{equation}
r_{E_0}
=
\sqrt{\langle r_{E_0}^{2}\rangle}.
\label{eq:GE0-rmsradius}
\end{equation}
Table~\ref{tab:radii} collects our results for the electric rms radii.
Since the form factors cannot be calculated directly at $Q^2=0$, the
radii are extracted by extrapolating the calculated form factors to the
origin using the Schlessinger point method (SPM), which has been widely
employed in hadron physics~\cite{yao:2021pyf,Xu:2019ilh,Binosi:2022ydc}.  This method
constructs continued-fraction approximants and, through statistical
sampling, provides the extrapolated value together with an estimate of
its uncertainty~\cite{Schlessinger:1968}. The same
extrapolation is, in principle, also required to determine
$G_{M1}(0)$.  In that case, however, the associated uncertainty is
numerically negligible at the precision quoted here and is therefore
not displayed separately.
 A comparison between the two triply-heavy baryons shows
that the electric radius of $\Omega_{ccc}^{++}$ is larger than that of
$\Omega_{bbb}^{-}$, demonstrating that the spatial extent decreases as
the heavy-quark mass increases. For $\Omega_{ccc}^{++}$, the available
lQCD estimate, obtained from a dipole reconstruction constrained by the
charge normalization and the single available nonzero-$Q^2$ point, is
in very good agreement with our result.

\subsection{Magnetic dipole form factor}
\label{sec:GM1}

Figure~\ref{fig:GM1-ccc-bbb} displays the normalized magnetic-dipole
form factors, $G_{M1}(Q^2)/G_{M1}(0)$, for
$\Omega_{ccc}^{++}$ and $\Omega_{bbb}^{-}$ in panels
(\textsf{A}) and (\textsf{B}), respectively.

\begin{figure*}[!t]
	\centering
	% ==================== 第一行 ====================
	\begin{minipage}[t]{0.49\textwidth}
		\centering

		\makebox[\linewidth][l]{%
			\textsf{\large A}%
			\hfill
			\textsf{\large$\Omega_{ccc}^{++}$}%
			\hfill
			\phantom{\textsf{\large A}}%
		}\\[-0.4em]

		\begin{minipage}[c]{1.8em}
			\centering
			\rotatebox{90}{%
				\textsf{$G_{E2}$, $G_{M3}$}
			}
		\end{minipage}
		\begin{minipage}[c]{0.9\textwidth}
			\centering
			\includegraphics[width=\linewidth]
			{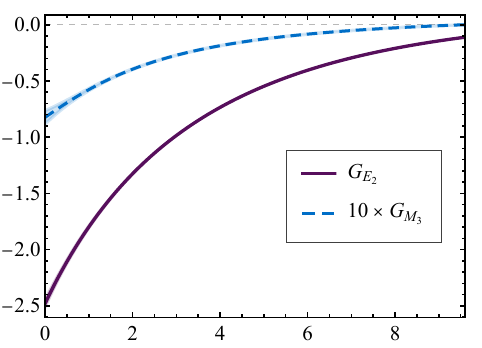}\\[-0.5em]
                			\textsf{$Q^2/\mathrm{GeV}^2$}
		\end{minipage}

	\end{minipage}
	\hfill
	\begin{minipage}[t]{0.49\textwidth}
		\centering

		\makebox[\linewidth][l]{%
			\textsf{\large B}%
			\hfill
			\textsf{\large$\Omega_{bbb}^{-}$}%
			\hfill
			\phantom{\textsf{\large B}}%
		}\\[-0.4em]

		\begin{minipage}[c]{1.8em}
			\centering
			\rotatebox{90}{%
				\textsf{$G_{E2}$, $G_{M3}$}
			}
		\end{minipage}
		\begin{minipage}[c]{0.9\textwidth}
			\centering
			\includegraphics[width=\linewidth]
			{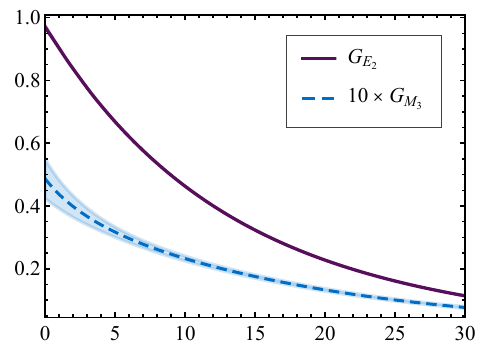}\\[-0.5em]
			\textsf{$Q^2/\mathrm{GeV}^2$}
		\end{minipage}

	\end{minipage}

\caption{Panels A and B show the electric quadrupole and magnetic octupole form factors of
\(\Omega_{ccc}^{++}\) and \(\Omega_{bbb}^{-}\) respectively.
The solid curve shows \(G_{E2}(Q^2)\), whereas the dashed curve
represents \(10\times G_{M3}(Q^2)\).  The band indicates the
uncertainty associated with the SPM extrapolation.}
\label{fig:GE2_GM3_bbb}
\end{figure*}
For $\Omega_{ccc}^{++}$, lQCD provides a result for $G_{M1}$ only at
the lowest nonzero momentum transfer,
$Q^2=0.183~\mathrm{GeV}^2$~\cite{Can:2015exa}.  This point is shown in
red in Fig.~\ref{fig:GM1-ccc-bbb}(\textsf{A}).  Since $G_{M1}(0)$
cannot be accessed directly in that calculation,
Ref.~\cite{Can:2015exa} assumes that, at low momentum transfer,
$G_{M1}$ has the same $Q^2$ dependence as $G_{E0}$.  The dashed
curve in panel (\textsf{A}) represents the corresponding dipole
reconstruction.  At low momentum transfer, our result is consistent
with the available lQCD point.  As $Q^2$ increases, our calculation
again becomes softer than the dipole reconstruction, in close analogy
with the behaviour observed for $G_{E0}$.

A comparison between panels (\textsf{A}) and (\textsf{B}) further
shows that the magnetic-dipole form factor of $\Omega_{bbb}^{-}$ is
harder than that of $\Omega_{ccc}^{++}$, indicating a more compact
magnetization distribution in the bottom sector.

Analogously to the electric case, the magnetic rms radius $r_{M_1}$ is
defined according to Eqs.~\eqref{eq:GE0-radius} and
\eqref{eq:GE0-rmsradius}, with $G_{E0}$ replaced by $G_{M1}$.
Our results are listed in Table~\ref{tab:radii}.  As in the electric
sector, the magnetic radius of $\Omega_{bbb}^{-}$ is smaller than that
of $\Omega_{ccc}^{++}$, consistent with the more compact structure of
the bottom system.  As noted above, the lQCD analysis assumes the same
low-$Q^2$ dependence for $G_{M1}$ and $G_{E0}$; within this
prescription, one therefore obtains $r_{M_1}=r_{E_0}$.

The magnetic moment of a spin-$3/2$ baryon is related to the magnetic-dipole
form factor at zero momentum transfer through
\begin{equation}
\mu_B
=
\frac{e}{2M_B}\,G_{M1}(0).
\label{eq:magnetic_moment}
\end{equation}
Table~\ref{tab:heavy-baryon-magnetic} compares our calculated magnetic
moments with predictions from other approaches.
For
$\Omega_{bbb}^{-}$, our result is close to the available model
predictions.  For $\Omega_{ccc}^{++}$, our magnetic moment is significantly larger
than the available lQCD determination, while remaining closer to most of other predictions.  It should be noted, however, that the lQCD calculation provides
$G_{M1}$ only at the lowest nonzero momentum transfer and assumes
that its low-$Q^2$ dependence follows that of $G_{E0}$ in order to
extrapolate to the origin. We therefore look forward to more precise lQCD determinations of
$G_{M1}(0)$.
Moreover, the reduction in the magnitude of the magnetic moment reflects
both the increase in the baryon mass and the evolution of $G_{M1}(0)$.
The latter is examined below in connection with the heavy-quark limit.

\subsection{Electric quadrupole and magnetic octupole form factors}

Figure~\ref{fig:GE2_GM3_bbb} displays the resulting
$G_{E2}(Q^2)$ and $G_{M3}(Q^2)$ for
$\Omega_{ccc}^{++}$ and $\Omega_{bbb}^{-}$ in panels
(\textsf{A}) and (\textsf{B}), respectively.  Both form factors
exhibit a smooth momentum dependence, while the larger uncertainty
associated with $G_{M3}$ reflects its stronger numerical sensitivity
at low $Q^2$.

The electric-quadrupole and magnetic-octupole form factors are obtained
from Eqs.~\eqref{eq:GE2_projection} and \eqref{eq:GM3_projection}.
Their numerical extraction becomes increasingly delicate in the
vicinity of $Q^2=0$, owing to the overall $1/Q^4$ and $1/Q^6$ factors
appearing in the projections of $G_{E2}$ and $G_{M3}$,
respectively.  Small numerical uncertainties in the projected
quantities $g_i$ are therefore amplified at low momentum transfer,
particularly for $G_{M3}$.  We consequently employ the SPM to
extrapolate the numerically stable results obtained at larger $Q^2$
toward $Q^2=0$.  The resulting values are:
% \begin{equation}
% G_{E2}^{\Omega_{ccc}^{++}}(0)=-2.48(4),
% \qquad
% G_{M3}^{\Omega_{ccc}^{++}}(0)=-0.0825(60),
% \end{equation}
% %
% and
% %
% \begin{equation}
% G_{E2}^{\Omega_{bbb}^{-}}(0)=0.971(8),
% \qquad
% G_{M3}^{\Omega_{bbb}^{-}}(0)=0.0487(64).
% \end{equation}
\begin{center}
\begin{tabular}{ccc}

&
$G_{E2}(0)$
&
$G_{M3}(0)$
\\
\hline
$\Omega_{ccc}^{++}$
&
$-2.48(4)$
&
$-0.0825(60)$
\\
$\Omega_{bbb}^{-}$
&
$0.971(8)$
&
$0.0487(64)$
\\
\end{tabular}
\end{center}
The opposite signs of the higher multipole form factors in the charm
and bottom sectors largely reflect the opposite electric charges of the
constituent quarks.  Their magnitudes, however, exhibit different
heavy-quark-mass dependences.  After normalizing by the corresponding baryon charge, this contrast becomes
even more apparent: $G_{E2}(0)/G_{E0}(0)$ decreases in magnitude
from $1.24$ to $0.971$, while $G_{M3}(0)/G_{E0}(0)$ remains unchanged within uncertainties.  Hence, the electric
quadrupole structure shows a clearer decreasing with increasing
heavy-quark mass, whereas the charge-normalized magnetic-octupole
structure is stable between the charm and bottom sectors.

\subsection{Electromagnetic properties in the heavy-quark limit}

The heavy-quark limit provides a useful reference for assessing
finite-mass and relativistic effects in triply-heavy baryons.  In the
following, we examine this evolution through two complementary
observables: the magnetic moment and the relation between the electric
and magnetic radii.

\subsubsection{Magnetic moment}

The heavy-quark reference value of the magnetic-dipole form factor can
be obtained from the leading NR magnetic-moment
operator.  
For a heavy quark of mass $m_Q$\footnote{In the DSE framework, the dressed-quark mass is momentum dependent, $M_Q(p^2)=B_Q(p^2)/A_Q(p^2)$. The mass $m_Q$
introduced here therefore serves only as the effective mass entering
the NR heavy-quark reference.} and electric charge $e_Q$, the corresponding operator is
\begin{equation}
\boldsymbol{\mu}_Q
=
\frac{e\,e_Q}{2\,m_Q}\,
\boldsymbol{\sigma}_Q,
\label{eq:heavy-quark-magnetic-moment-operator}
\end{equation}
where $e_Q$ denotes the quark charge in units of the elementary charge
$e$.  For the ground-state $\Omega_{QQQ}$ baryon, the spins of the
three identical quarks are aligned to form a total spin $J=3/2$.
Consequently, in the maximally polarized state $J_z=3/2$, the NR
magnetic moment is
\begin{equation}
\mu_{\Omega_{QQQ}}^{\rm NR}
=
3\,\frac{e\,e_Q}{2\,m_Q}.
\label{eq:heavy-quark-magnetic-moment}
\end{equation}
On the other hand, the magnetic moment of a spin-$3/2$ baryon is
related to its magnetic-dipole form factor through
\begin{equation}
\mu_{\Omega_{QQQ}}
=
\frac{e}{2M}\,
G_{M1}(0),
\end{equation}
where $M$ is the baryon mass.  Combining the two expressions gives
\begin{equation}
G_{M1}^{\rm NR}(0)
=
3\,e_Q\,\frac{M}{m_Q}.
\label{eq:GM1-NR}
\end{equation}
In the strict heavy-quark limit, $M\to 3m_Q$, and hence
\begin{equation}
G_{M1}^{\rm NR}(0)
=
9\,e_Q.
\label{eq:GM1-HQ}
\end{equation}
Therefore,
\begin{equation}
G_{M1}^{\Omega_{ccc}^{++}}(0)
\longrightarrow 6,
\qquad
G_{M1}^{\Omega_{bbb}^{-}}(0)
\longrightarrow -3.
\label{eq:GM1-heavy-quark-reference}
\end{equation}

To quantify the approach to these reference values, we define
\begin{equation}
\mathcal{R}_{M_1}^{(Q)}
=
\frac{G_{M1}^{\Omega_{QQQ}}(0)}
     {G_{M1}^{\rm NR}(0)},
\qquad
\Delta_{M_1}^{(Q)}
=
\left|1-\mathcal{R}_{M_1}^{(Q)}\right|.
\label{eq:gm1-heavy-quark-ratio}
\end{equation}
Using our calculated values,
$G_{M1}^{\Omega_{ccc}^{++}}(0)=6.40$ and
$G_{M1}^{\Omega_{bbb}^{-}}(0)=-3.14$, we obtain
\begin{equation}
\mathcal{R}_{M_1}^{(c)}
=
\frac{6.40}{6}
=
1.067,
\qquad
\mathcal{R}_{M_1}^{(b)}
=
\frac{-3.14}{-3}
=
1.047.
\label{eq:gm1-heavy-quark-ratios-results}
\end{equation}
The corresponding deviations are
\begin{equation}
\Delta_{M_1}^{(c)}
=
0.067,
\qquad
\Delta_{M_1}^{(b)}
=
0.047.
\label{eq:gm1-heavy-quark-deviations}
\end{equation}
Hence, the deviation decreases from approximately $6.7\%$ for the
$\Omega_{ccc}^{++}$ to $4.7\%$ for the $\Omega_{bbb}^{-}$,
corresponding to a reduction of about $30\%$.  The value of
$G_{M1}(0)$ for the $\Omega_{bbb}^{-}$ therefore lies closer to its
heavy-quark-limit reference value, exhibiting the expected approach
toward heavy-quark behaviour as the quark mass increases.

\subsubsection{Electric and magnetic radii}

A complementary perspective is provided by comparing the electric and
magnetic distributions.  In the leading NR picture, the three identical and spin-aligned heavy quarks occupy the
same spatial state.  Within the leading one-body description, the
electric and magnetic distributions are therefore governed by the same
underlying spatial wave function~\cite{Buchmann:1996,Kaskulov:2003}.  Consequently, in the
NR heavy-quark limit one expects
\begin{equation}
\left\langle r_{E_0}^{2}\right\rangle
\longrightarrow
\left\langle r_{M_1}^{2}\right\rangle,
\qquad
r_{E_0}\longrightarrow r_{M_1}.
\label{eq:nr-electric-magnetic-radii}
\end{equation}
Deviations from this relation therefore provide a measure of
finite-mass and relativistic effects in the electromagnetic structure
of the triply-heavy baryons.

Following the same strategy as above, we define
\begin{equation}
\mathcal{R}_{r}^{(Q)}
=
\frac{r_{M_1}^{(Q)}}{r_{E_0}^{(Q)}},
\qquad
\Delta_{r}^{(Q)}
=
\left|1-\mathcal{R}_{r}^{(Q)}\right|.
\label{eq:electric-magnetic-radius-ratio}
\end{equation}
%
% Thus,
% $\mathcal{R}_{r}^{(Q)}\to1$ and
% $\Delta_{r}^{(Q)}\to0$ in the heavy-quark
% limit.
Using the central values of the radii listed in Table~\ref{tab:radii}, one finds
\begin{align}
\mathcal{R}_{r}^{(c)}
&=
\frac{0.279}{0.287}
=
0.972,
&
\Delta_{r}^{(c)}
&=
0.028,
\nonumber\\
\mathcal{R}_{r}^{(b)}
&=
\frac{0.138}{0.140}
=
0.986,
&
\Delta_{r}^{(b)}
&=
0.014.
\label{eq:electric-magnetic-radius-results}
\end{align}
Hence, the difference between the electric and magnetic radii is reduced
by approximately a factor of two when going from
$\Omega_{ccc}^{++}$ to $\Omega_{bbb}^{-}$.  At the same time, the
absolute radii decrease substantially, reflecting the increasingly
compact spatial structure of the baryon.  The approach
$r_{M_1}/r_{E_0}\to1$ therefore indicates
that the electric and magnetic distributions become increasingly
similar as the quark mass increases, consistent with the gradual
approach toward the leading NR heavy-quark limit. This trend has also been found in meson studies
within the same DSE framework, where the spatial distributions become
more compact and differences between related radii are reduced as the
quark mass increases~\cite{Xu:2024vkn,Xu:2026rtu}.

\section{Conclusions and outlook}
\label{sec:concl}
In this work, we have investigated the ground-state
$\Omega_{ccc}^{++}$ and $\Omega_{bbb}^{-}$ baryons within a
Poincar\'e-covariant three-body Faddeev framework.  The dressed
heavy-quark propagators, baryon Faddeev amplitudes, and electromagnetic
current are treated consistently within the rainbow--ladder truncation.
With a common symmetry-preserving interaction, we obtain a good
description of the relevant heavy-meson and triply-heavy-baryon masses
and calculate the complete set of elastic spin-$3/2$ electromagnetic
multipole form factors,
$G_{E0}$, $G_{M1}$, $G_{E2}$, and $G_{M3}$.

For $\Omega_{ccc}^{++}$, our low-$Q^2$ electric form factor and electric
radius are in good agreement with the available lQCD results.
At larger momentum transfer, our direct result becomes softer than the
dipole reconstruction based on the limited lQCD data, illustrating the uncertainty associated with extrapolating simple pole-like parametrizations beyond the constrained momentum region. The
calculated magnetic moments,
$\mu_{\Omega_{ccc}^{++}}=1.26\,\mu_N$ and
$\mu_{\Omega_{bbb}^{-}}=-0.207\,\mu_N$, are broadly consistent with
existing model expectations, although the $\Omega_{ccc}^{++}$ result
is somewhat larger than the available lQCD determination.  We also
determine the electric and magnetic rms radii of both systems.  In each
case, the corresponding radius of $\Omega_{ccc}^{++}$ is larger than
that of $\Omega_{bbb}^{-}$.  Consistently, the leading electromagnetic
form factors, $G_{E0}$ and $G_{M1}$, are harder for
$\Omega_{bbb}^{-}$ than for $\Omega_{ccc}^{++}$.  These results show
that both the electric-charge and magnetization distributions become
more compact as the heavy-quark mass increases from charm to bottom.

For the higher multipole form factors, $G_{E2}$ and $G_{M3}$, we
employ the SPM to extrapolate the numerically stable finite-$Q^2$
results to the origin.  We obtain
$G_{E2}^{\Omega_{ccc}^{++}}(0)=-2.48(4)$ and
$G_{M3}^{\Omega_{ccc}^{++}}(0)=-0.0825(60)$, compared with
$G_{E2}^{\Omega_{bbb}^{-}}(0)=0.971(8)$ and
$G_{M3}^{\Omega_{bbb}^{-}}(0)=0.0487(64)$.
The electric-quadrupole structure exhibits a more pronounced
charm-to-bottom evolution, whereas the charge-normalized
magnetic-octupole structure remains comparatively stable.

The comparison of the charm and bottom systems provides an additional
perspective on the approach toward the heavy-quark limit.  For three
identical spin-aligned heavy quarks, the leading nonrelativistic limit
gives $G_{M1}(0)\to 3\,Q_{\Omega_{QQQ}}$.  The deviation of our result
from this reference decreases from $6.7\%$ for $\Omega_{ccc}^{++}$ to
$4.7\%$ for $\Omega_{bbb}^{-}$.  Independently, the ratio of magnetic
to electric radii evolves from $r_{M_1}/r_{E_0}=0.972$ in the charm
sector to $0.986$ in the bottom sector, approaching the equality
expected in the NR limit.

A natural extension of the present work is to use the complete set of
electromagnetic form factors to construct light-front transverse charge
and magnetization densities.  Such densities provide a spatial
representation of electromagnetic structure in the transverse plane
and have recently been investigated for the nucleon~\cite{Xu:2025ups,Lorce:2020onh}.  Extending this analysis to the $\Delta$,
$\Omega^-$, $\Omega_{ccc}^{++}$, and $\Omega_{bbb}^{-}$ would
provide a unified view of the flavour evolution of spin-$3/2$ baryon
structure from the light and strange sectors through charm and bottom.
% Further improvements beyond rainbow--ladder, including ACM and other
% nonleading structures of the quark--gluon vertex, would also provide a
% useful assessment of truncation effects, particularly for the
% spin-sensitive higher multipoles.

\section*{Acknowledgments}
Zhao-Qian Yao acknowledges support from Helmholtz-Zentrum
Dresden-Rossendorf through the High Potential Programme and
thanks ECT* for its hospitality and support during the workshop
``Theoretical Motivations and Empirical Evidence for Diquark
Correlations in Hadrons'', where part of this work was developed.

\appendix

\section{}
\label{sec:extraction}
To extract the four independent form factors from the calculated
current in Eq.~\eqref{eq:spin32_current}, we contract
$\mathcal{J}^{\mu,mn}$ with four linearly independent projection
tensors,
\begin{align}
\mathcal{T}_{1}^{\mu,mn}
&=
\widehat P^\mu
\widehat P^m
\widehat P^n,
&
\mathcal{T}_{2}^{\mu,mn}
&=
\widehat P^\mu
\delta^{mn},
\nonumber\\
\mathcal{T}_{3}^{\mu,mn}
&=
\gamma^\mu
\widehat P^m
\widehat P^n,
&
\mathcal{T}_{4}^{\mu,mn}
&=
\gamma^\mu
\delta^{mn}.
\label{eq:projection4}
\end{align}
The corresponding scalar projections are defined by
\begin{equation}
g_i
=
\operatorname{Tr}_{D}
\left[
 \mathcal{T}_{i}^{\mu,mn}
 \mathcal{J}^{\mu,mn}
\right],
\qquad i=1,\ldots,4.
\label{eq:spin32_scalar_projections}
\end{equation}
Substitution of the current decomposition,
Eq.~\eqref{eq:spin32_current}, into
Eq.~\eqref{eq:spin32_scalar_projections} produces four linearly
independent equations for the four invariant form factors
$F_{1,2,3,4}(Q^2)$. Solving this linear system yields the invariant
form factors, from which the physical multipole form factors
$G_{E0}$, $G_{M1}$, $G_{E2}$ and $G_{M3}$ follow through
Eqs.~\eqref{eq:GE0_def}--\eqref{eq:GM3_def}.
Equivalently, the physical multipole form factors can be expressed
directly in terms of the four scalar projections $g_i$:
\begin{align}
G_{E0}(Q^2)
&=
\frac{i\left(2g_1-g_2\right)}
     {4\sqrt{1+\tau}},
\label{eq:GE0_projection}
\\[1mm]
G_{M1}(Q^2)
&=
\frac{9iM^2}
     {10Q^2}
\left[
\frac{2g_1-g_2}{\sqrt{1+\tau}}+
\left(-2g_3+g_4\right)
\right],
\label{eq:GM1_projection}
\\[1mm]
G_{E2}(Q^2)
&=
-\frac{3iM^2}
       {2Q^4  \sqrt{1+\tau}}
\left[
-g_2Q^2+2g_1\left(6M^2+Q^2\right)
\right],
\label{eq:GE2_projection}
\\[1mm]
G_{M3}(Q^2)
&=
\frac{3iM^4}
     {Q^6\sqrt{1+\tau}}
\Bigg[
g_2Q^2
-2g_1\left(5M^2+Q^2\right)
\nonumber\\
&+\left(
-g_4Q^2
+2g_3\left(5M^2+Q^2\right)
\right)
\sqrt{1+\tau}
\Bigg],
\label{eq:GM3_projection}
\end{align}
where $\tau=Q^2/(4M^2)$. These relations allow the four multipole
form factors to be extracted directly from the numerically calculated
scalar projections, without first solving explicitly for the
invariant form factors $F_i(Q^2)$.
Notably, the first two projectors in Eqs.~\eqref{eq:projection4} have the same tensor structure as those
commonly employed in previous studies~\cite{Nozawa:1990,
Nicmorus:2010,SanchisAlepuz:2013}. In the last two projectors, however, we use $\gamma^\mu$ instead of the transverse matrix
$\gamma_T^\mu$. This alternative choice remains linearly independent hence
leads to the same physical form factors. It has the practical
advantage of producing considerably shorter algebraic expressions,
thereby reducing the cost of the symbolic manipulations and
accelerating the subsequent numerical evaluation.

\bibliographystyle{apsrev4-2}
\bibliography{bibliography}

\end{document}